\documentclass[twocolumn,amsmath,amssymb,superscriptaddress,prl,a4paper]{revtex4}
\usepackage{graphicx}
\usepackage[paperwidth=212mm,paperheight=297mm,centering,hmargin=1.65cm,vmargin=2.6cm]{geometry}

\usepackage{graphicx}                       
\usepackage{color}                          
\usepackage{hyperref}                       
\usepackage{dsfont}
\usepackage{mathrsfs}
\usepackage{bigints}
\usepackage{bbold}
\usepackage{amsthm}
\usepackage{setspace}
\usepackage{epstopdf}

\newcommand{\beq}{\begin{eqnarray}}
\newcommand{\eeq}{\end{eqnarray}}

\DeclareGraphicsExtensions{.jpg,.mps,.pdf,.png,.gif}

\begin{document}


%
%
\author{Antoine Reigue}

\affiliation{Sorbonne Universit\'es, UPMC Univ Paris 06, CNRS UMR 7588, Institut des NanoSciences de Paris, F-75005, Paris, France}





\author{Aristide Lema\^{i}tre}

\affiliation{Centre de Nanosciences et de Nanotechnologies, CNRS, Univ. Paris-Sud, Universit\'e Paris-Saclay, C2N-Marcoussis, 91460 Marcoussis, France}

\author{Carmen Gomez Carbonell}

\affiliation{Centre de Nanosciences et de Nanotechnologies, CNRS, Univ. Paris-Sud, Universit\'e Paris-Saclay, C2N-Marcoussis, 91460 Marcoussis, France}

\author{Christian Ulysse}

\affiliation{Centre de Nanosciences et de Nanotechnologies, CNRS, Univ. Paris-Sud, Universit\'e Paris-Saclay, C2N-Marcoussis, 91460 Marcoussis, France}

\author{Kamel Merghem}

\affiliation{Centre de Nanosciences et de Nanotechnologies, CNRS, Univ. Paris-Sud, Universit\'e Paris-Saclay, C2N-Marcoussis, 91460 Marcoussis, France}

\author{St\'ephane Guilet}

\affiliation{Centre de Nanosciences et de Nanotechnologies, CNRS, Univ. Paris-Sud, Universit\'e Paris-Saclay, C2N-Marcoussis, 91460 Marcoussis, France}

\author{Richard Hostein}

\affiliation{Sorbonne Universit\'es, UPMC Univ Paris 06, CNRS UMR 7588, Institut des NanoSciences de Paris, F-75005, Paris, France}

\author{Valia Voliotis}
\email{voliotis@insp.jussieu.fr}

\affiliation{Sorbonne Universit\'es, UPMC Univ Paris 06, CNRS UMR 7588, Institut des NanoSciences de Paris, F-75005, Paris, France}

\title{Resonance fluorescence revival in a voltage-controlled semiconductor quantum dot}

\begin{abstract}

We demonstrate systematic resonance fluorescence recovery with near-unity emission efficiency in single quantum dots embedded in a charge-tunable device in a wave-guiding geometry.  The quantum dot charge state is controlled by a gate voltage, through carrier tunneling from a close-lying Fermi sea,  stabilizing the resonantly photocreated electron-hole pair. The electric field cancels out the charging/discharging mechanisms from nearby traps toward the quantum dots, responsible for the usually observed inhibition of the resonant fluorescence. Fourier transform spectroscopy as a function of the applied voltage shows a strong increase of the coherence time though not reaching the radiative limit. These charge controlled quantum dots act as quasi-perfect deterministic single-photon emitters, with one laser pulse converted into one emitted single photon.

\end{abstract}

\maketitle

Semiconductor quantum dots (QDs) are commonly considered as artificial atoms due to their discrete electronic structure and are very efficient sources of single and indistinguishable photons~\cite{santori2002indistinguishable, somaschi2016nearoptimal}. Their potential for applications is high as they can be easily integrated into nanophotonic devices~\cite{makhonin2014waveguide, buckley2012engineered}, defining building blocks for quantum information processing in the solid-state~\cite{gao2015coherent, lodahl2015interfacing}. 
During the past decade, a lot of effort has been devoted to minimize dephasing processes due to the coupling of QDs to their surrounding solid-state matrix. Indeed, coupling to phonons~\cite{reigue2017probing, grange2017reducing} as well as time jitter~\cite{unsleber2015twophotons} reduces the degree of indistinguishability and charge or spin noise~\cite{berthelot2006unconventional, houel2012probing, kuhlmann2013charge} lead to inhomogeneous broadening of the emission line. Therefore, strictly resonant excitation of the QD s-shell has appeared as an essential ingredient~\cite{muller2007resonance, melet2008resonant} to preserve the coherence properties of the emitted photons.

Charge noise is detrimental as it strongly limits or even suppresses the QD resonance fluorescence (RF) \cite{reinhard2012strongly, nguyen2012optically}.  The RF quench has been attributed to the structure residual doping and defects which create a fluctuating electrostatic environment. This can lead to a Coulomb blockade effect preventing the photocreation of an electron-hole pair in the QD \cite{nguyen2012optically, nguyen2013photoneutralization}. To circumvent this difficulty and recover the RF, an additional very low power non-resonant laser can be used. 
Although this technique has been succesful \cite{gazzano2013bright, monniello2014indistinguishable, chen2016characterization}, the exact  physical process of this non-resonant pump has not been sufficiently addressed \cite{chen2016characterization}.  Moreover, the lack of control of this non-resonant pump prevents the realization of a fully on-demand single-photon source with a high degree of coherence and indistinguishability.


Here, we show how RF quenching can be avoided by a suitably designed voltage-controlled device. The resonant excitation is realized in an in-plane waveguide geometry~\cite{reigue2017probing}, while the single photons are collected from the top. This geometry yields an almost complete suppression of the laser scattered light on the RF detection side. By controlling the QD electric field environment by a gate voltage, the charging/discharging mechanisms from the nearby trap states to the QD are disabled. The resonantly photocreated electron-hole pairs give rise to a very intense RF line and an increase of the coherence time. However the radiative limit is not reached suggesting that charge and/or spin noise are still present in the structure leading to residual inhomogeneous broadening. Our device appears, nonetheless, as an almost perfect single-photon source suitable for quantum technology applications.

\begin{figure*}[htb]
\begin{center}
\includegraphics[width=17cm]{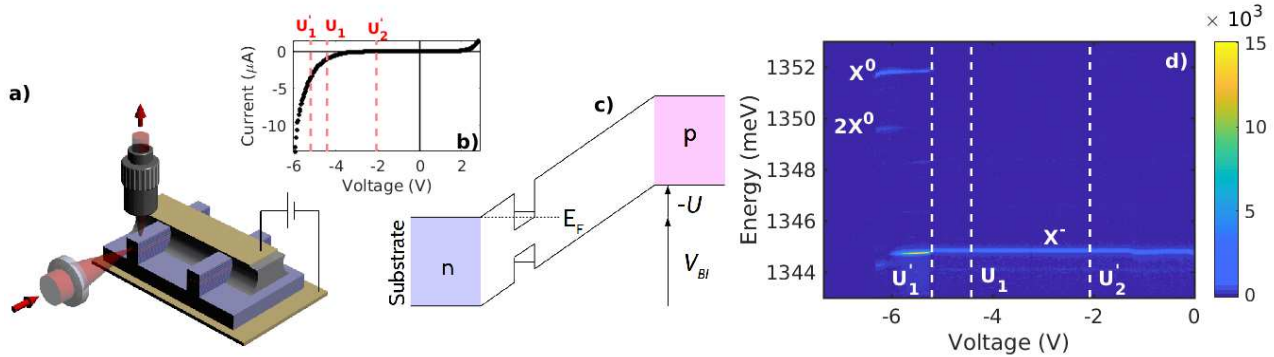}
\caption{(a) Schematic structure of the sample: etched ridges (in blue) containing the QDs and microscope objectives for excitation and detection. The electrical contacts are represented in yellow. (b) Typical I-V curve at 4 K indicating the different threshold voltages. (c) Band structure of the sample. (d) Typical micro-photoluminescence map of a QD as a function of the applied gate voltage for non-resonant excitation. $U'_1$ is the threshold for the emergence of the $X^0$ transition. $U_1$ and $U'_2$ indicate the region where the $X^-$ RF exists (see text). }
\label{experimental_setup}
\end{center}
\end{figure*}

A low density InAs/GaAs self-assembled QD layer, grown by molecular beam epitaxy, was embedded at the center of a \textit{p-i-n} doped GaAs/AlAs microcavity on a n-doped GaAs (001) substrate. The Bragg mirrors, consisting of  12 p-doped pairs on the top and 24 n-doped pairs at the bottom, were designed to maximise the luminescence collection efficiency. The quality factor is only a few hundreds and does not induce any significant Purcell effect.  Deep (approximatively $1.5 \, \mu \mathrm{m}$) ridges were etched by inductively coupled plasma etching realizing one-dimensional waveguides with 0.8 to $1.2 \, \mu\mathrm{m}$ width. Standard ohmic contacts were deposited and annealed on the back side. Top contacts were realized by resist planarization and deposition of Ti/Au stripes perpendicular to ridges. The resist was then etched away between the stripes. A schematic view of the experimental geometry, a characteristic I-V curve and the simplified band structure are depicted in Fig.~\ref{experimental_setup} (a), (b) and (c) respectively.
In the following, $U$ will denote the external applied bias and $V_{\mathrm{BI}}$ the built-in voltage. Then, the QD potential is $(V_{\mathrm{BI}} - U)$ for reverse bias. 

To investigate the resonant and non-resonant fluorescence, the sample and the microscope objectives are mounted inside a He closed cycle temperature-variable cryostat.
For the  excitation, a tunable picosecond Ti-Sapphire laser beam is focused on a ridge cleaved edge. The pulse propagates along the one-dimensional waveguide and pumps one or several QDs. The luminescence of a single of these dots is collected from the ridge top surface using confocal microscopy. The  signal is coupled to a monomode optical fiber and sent to the detection setup composed by a spectrometer and a CCD camera with spectral resolution of 60 $\mathrm{\mu eV}$. It can also be sent either to a Michelson interferometer for Fourier transform spectroscopy, or to a photon correlation setup~\cite{reigue2017probing}.

Ten different QDs were studied from different ridges and the same behavior was observed for all of them. In the following we show results obtained from the same QD. We present in Fig.~\ref{experimental_setup}(d) a typical micro-photoluminescence ($\mu$-PL) map under non-resonant excitation (at the wetting layer energy), as a function of the applied bias voltage. Polarization and power-dependent measurements have been performed to identify unambiguously the QD neutral electron-hole pair $X^0$ and charged $X^-$ states as well as the double electron-hole pair state $2X^0$ (Fig.~\ref{experimental_setup}(d)). The  $X^0$ emission threshold is around $U'_1 = -5.2 \, \mathrm{V}$. From the energy positions of the $X^0$ and $X^-$ lines in the $\mu \mathrm{PL}$ map, we find a binding energy for the $X^-$ charged complex of about $7 \, \mathrm{meV}$ in agreement with several reported experimental observations~\cite{warburton2000optical, smith2003carrier, bennett2010giant}.

\begin{figure}[!h]
\begin{center}
\includegraphics[width=8cm]{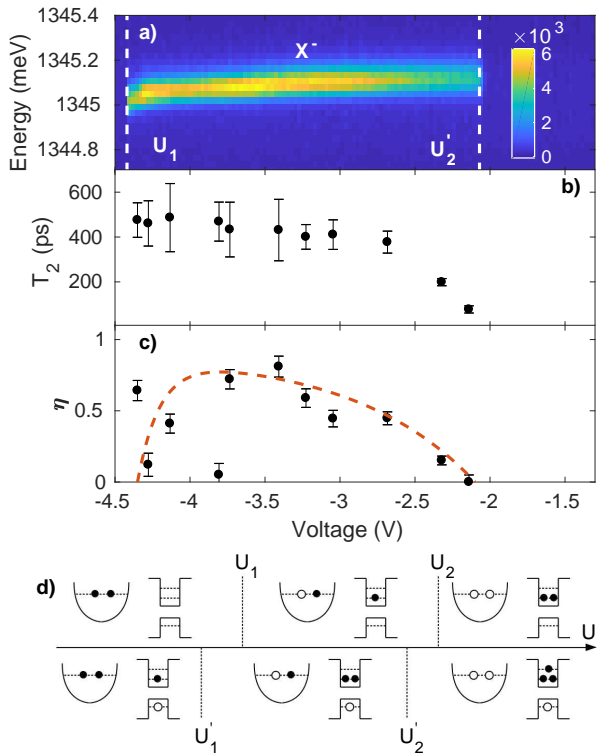}
\end{center}
\caption{RF experiments on the $X^-$ line as a function of the gate voltage: (a) $\mu$-PL map under $\pi$-pulsed excitation. (b) Coherence time $T_2$ measured by Fourier transform spectroscopy. (c) Inhomogeneous contribution of the pseudo-Voigt profile (see text). The red dotted curve is a guide to the eye. (d) Diagram of the different QD electronic states with applied bias where one or two electrons can tunnel from the Fermi sea to the dot. The upper (lower) panel shows the configurations without (with one) electron-hole pair photocreated in the QD. The black and open dots correspond to electrons  and holes respectively. }
\label{excitation_resonante}
\end{figure}

RF experiments have been performed on the $X^-$ line for all the studied QDs. Fig.~\ref{excitation_resonante}(a) shows the $\mu$-PL map under resonant $\pi$-pulsed excitation. We clearly observe two thresholds: at $U'_2 = -2.1 \, \mathrm{V}$ corresponding to the appearance of the $X^-$ RF and at  $U_1 = -4.4 \, \mathrm{V}$ where the RF vanishes~\footnote{It is worth noticing that for non-resonant excitation, electrons can be captured in the QD, and emission from the $X^{-}$ state can be observed beyond the expected thresholds.}. We represent schematically in Fig.~\ref{excitation_resonante}(d) the different electronic states for an empty QD (upper panel) and a QD containing a photo-created electron-hole pair (lower panel) as a function of the applied gate voltage. The threshold $U_1$ ($U_2$) corresponds to the bias voltage where it is energetically favorable for one (two) electron(s) to tunnel from the Fermi sea into the QD in the absence of electron-hole pair.
When an electron-hole pair is photocreated in the dot filled with one electron, the threshold $U_1$ is shifted towards a lower gate voltage $U'_1$. Indeed, it is straightforward to show that ($U_{1} - U'_1$) has the same sign as the energy difference between  $X^0$ and $X^{-}$ which is positive. This is due to the stronger hole localization in usual InAs QDs~\cite{bester2003compositional, gong2008electronic, andrade2016manybody} which bounds the $X^-$ complex. In the same way, ($U_{2} - U'_2$) has the same sign as the energy difference between  $X^-$ and $X^{2-}$.  This quantity has been measured to be positive \cite {warburton2000optical, kleemans2010manybody, smith2003carrier, baier2001optical}, thus the threshold $U_2$ is also shifted toward a lower gate voltage $U'_2$.

\begin{figure}[h]
\begin{center}
\includegraphics[width=8cm]{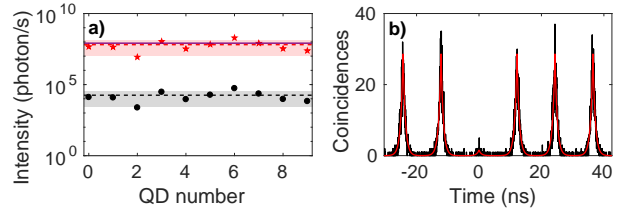}
\end{center}
\caption{(a) RF intensity (number of photons per second in log scale) for all the studied QDs: raw data are in black dots and the corrected data by the collection rate of the experimental setup are shown in red stars. The shaded areas represent the data standard deviation and the dotted line corresponds to the mean value. The solid line is the number of laser pulses per second. (b) Coincidences histogram obtained in an HBT setup, measuring a very low multi-photon emission probability $g^{(2)}_{\mathrm{HBT}} = 0.05 \pm 0.02$.}
\label{HBT_efficiency}
\end{figure}

Therefore, the QDs states with and without one photocreated electron-hole pair, can be deterministically controlled at low temperature through the gate voltage, and RF recovery is expected for any desired transition. Moreover, the RF intensity measurements after taking into account the collection efficiency of our optical setup, i.e. collection at the first objective and losses in the optical path, show the high efficiency of the gate control. The corrected number of  photons per second is close to the laser pulse repetition rate ($82 \, \mathrm{MHz}$), indicating an almost one-to-one conversion of one laser pulse into one single photon (Fig.~\ref{HBT_efficiency}(a)). This shows the great potential of such a structure for the developement of high efficiency single photon source~\cite{somaschi2016nearoptimal}.

 The RF revival is assisted by an increase of the coherence time $T_2$ as measured by Fourier transform spectroscopy (FTS) versus the applied bias (Fig.~\ref{excitation_resonante}(b)). In the radiative limit where a QD is only coupled to the electromagnetic field, the RF line is \textit{homogeneously} broadened and $T_2 = 2T_1$. However, a QD is an open quantum system strongly coupled to its environment. For this reason, the RF line is also \textit{inhomogeneously} broadened by the interaction with the phonon bath~\cite{besombes2001acoustic}, the fluctuating charges in the vicinity of the QD~\cite{berthelot2006unconventional, houel2012probing} and the nuclear spins~\cite{kuhlmann2013charge}. In the presence of a close-lying charge reservoir, the additional dephasing process due to the interaction between the QD discrete excited state and the Fermi sea continuum gives rise to an \textit{homogeneous} broadening of the RF line~\cite{cohen2001processus_complement_CI_anglais}. Therefore, the value of $T_2$ and the shape of the RF will depend on the competition between these different dephasing processes.

The FTS interference contrast is ajusted by a pseudo-Voigt profile as detailed in~\cite{reigue2017probing}. Fig.~\ref{excitation_resonante}(b) shows a rapid increase of the coherence time close to the threshold $U \approx U'_2$ followed by a plateau where $T_2 \lesssim T_1$. $T_1 = 580 \pm 5 \, \mathrm{ps}$ is the resonant radiative lifetime and is independent on the bias voltage. The pseudo-Voigt profile is composed partly by the inhomogeneous contribution discussed above, hereafter called $\eta$ and shown in Fig.~\ref{excitation_resonante}(c). $\eta$ reaches a maximum value of 0.8 around $-3.5 \, \mathrm{V}$ and decreases to zero when the bias voltage gets close to the thresholds $U_1$ and $U'_2$. $\eta=0$ corresponds to an homogeneously broadened RF line, while $\eta =1$ indicates that inhomogeneous broadening is dominant.

We now discuss the behavior of $T_2$ and $\eta$ as a function of the gate voltage.
For $U \approx U'_2$ where an electron from the Fermi sea can easily tunnel  into and outside the dot, the coupling between the reservoir and the QD is strong. Therefore, the homogeneous contribution is dominant, $\eta$ is almost zero and the value of $T_2$ goes to zero. Between $U_1$ and $U'_2$, the coupling with the Fermi sea is weak, the electron being stabilized in the dot, and the coherence time increases. However, at the same time, the interaction with the solid matrix is still present explaining the high value of $\eta$. At $4 \, \mathrm{K}$, the loss of coherence due to the interaction with the phonon bath can be estimated~\cite{bylander2003interference, iles2017phonon} and the maximum expected value of $T_2$ would be about 1 ns. However, the experimental value is lower, suggesting that charge and spin noise are still present and responsible for the residual loss of coherence~\cite{alKhuzheyri2016resonance}.
For $U \approx U_1$, where the decoherence processes are again dominated by the interaction with the Fermi sea, we observe a decrease of the inhomogeneous contribution but no variation of the coherence time.
Indeed, the thresholds for $X^0 / X^-$ ($U_1$) and $X^- / X^{2-}$ ($U'_2$) do not obey to the same dynamics with applied bias, as also discussed in Ref.~\cite{kurzmann2016optical}. The $X^{2-}$ state has an electron in the p-level with a more delocalized wave function compared to the $X^-$ state where the electrons lie on the s-level. In the former case, the interaction with the Fermi reservoir is stronger and occurs on a larger voltage range, as observed in the variation of the RF intensity (Fig.~\ref{excitation_resonante}(a)). On the contrary, close to $U_1$, the RF intensity drops too fast with the gate voltage so that expected decrease of $T_2$ cannot be observed.

Finally, we performed second order correlation measurements to characterize the single-photon emission purity. Fig.~\ref{HBT_efficiency}(b) presents the coincidences histogram obtained under $\pi$-pulsed resonant excitation. We used a Hanbury-Brown-Twiss (HBT) setup with an integration time of $200 \, \mathrm{s}$. We clearly observe an antibunching at zero delay, corresponding to a very low multiphoton emission probability. A multi-exponential decay fit (red solid line) was used to extract the value of $g^{(2)}_{\mathrm{HBT}} = 0.05 \pm 0.02$. We attribute the correlations at zero delay to the scattered laser which nonetheless remains very low thanks to the sophisticated sample structure.

In summary, we present an optimized voltage-controlled device allowing the deterministic control of the QD charge state and the consistent revival of the RF with near unity efficiency. Additionally, strictly resonant coherence time measurements evidence the strong interaction between the QDs and the Fermi sea close to the thresholds with one, or two electrons in the dot. 
To go further in the understanding and modeling of the coherence time behavior, it would be interesting to treat the coupling between the Fermi sea and the QD with an Anderson-like model~\cite{andrade2016manybody, anderson1967infrared}. Using this model, interesting experimental features have been addressed like the observation of Mahan excitons~\cite{kleemans2010manybody}, the hybridization of the QD states with a filled continuum~\cite{dalgarno2008optically}, or the emergence of Kondo correlations~\cite{haupt2013nonequilibrium}. 
Within the voltage range where the RF is stabilized, minimizing charge and spin noise is still an issue. Additional experiments, for instance, under external applied magnetic field \cite{malein2016screening} or coherent population trapping~\cite{ethier2017improving} would allow polarizing the nuclear spin bath thus reducing the fluctuating spin environment.  The contribution of each dephasing mechanism could then be unambiguously assigned.
Finally, second order correlation measurements show very low multiphoton emission probability making this kind of device suitable for quantum information applications.

\begin{acknowledgements}
The authors acknowledge
Beno\^it Eble, Paola Atkinson, Monique Combescot and Fran\c{c}ois Dubin for helpful discussions. 

\end{acknowledgements}


\bibliographystyle{apsrev4-1}
\bibliography{/home/antoine/Bureau/These/Redaction_these/These_antoine_corrections/biblio}







\end{document}